\documentstyle[12pt,epsf]{article}
\def\frac#1#2{{#1\over#2}}

\font\titlefont=cmcsc10 scaled \magstep2
\begin{document}

\centerline {\titlefont Virtual Black Holes   }
\vskip .9truein
\centerline {S.~W.~Hawking} \vskip .5truein
\centerline {Department of Applied
Mathematics and Theoretical Physics}
\centerline {University of Cambridge}
\centerline {Silver Street}
\centerline {Cambridge CB3 9EW}
\centerline {UK}
\vskip .9truein
\centerline {\bf Abstract}
 \bigskip

One would expect spacetime to have a foam-like structure on the Planck
scale with a very high topology. If spacetime is simply connected
(which is assumed in this paper), the non-trivial homology occurs in
dimension two, and spacetime can be regarded as being essentially the
topological sum of $S^2\times S^2$ and $K3$ bubbles. Comparison with
the instantons for pair creation of black holes shows that the
$S^2\times S^2$ bubbles can be interpreted as closed loops of virtual
black holes. It is shown that scattering in such topological
fluctuations leads to loss of quantum coherence, or in other words, to
a superscattering matrix $\$ $ that does not factorise into an $S$
matrix and its adjoint. This loss of quantum coherence is very small
at low energies for everything except scalar fields, leading to the
prediction that we may never observe the Higgs particle. Another
possible observational consequence may be that the $\theta $ angle of
QCD is zero without having to invoke the problematical existence of a
light axion. The picture of virtual black holes given here also
suggests that macroscopic black holes will evaporate down to the
Planck size and then disappear in the sea of virtual black holes.

\vfil \eject

\section{Introduction}

It was John Wheeler who first pointed out that quantum fluctuations in
the metric should be of order one at the Planck length. This would
give spacetime a foam-like structure that looked smooth on scales
large compared to the Planck length. One might expect this spacetime
foam to have a very complicated structure, with an involved
topology. Indeed, whether spacetime has a manifold structure on these
scales is open to question. It might be a fractal. But manifolds are
what we know how to deal with, whereas we have no idea how to formulate
physical laws on a fractal. In this this paper I shall therefore
consider how one might describe spacetime foam in terms of manifolds
of high topology.

I shall take the dimension of spacetime to be four. This may sound
rather conventional and restricted, but there seem to be severe
problems of instability with Kaluza Klein theories. There is something
rather special about four dimensional manifolds, so maybe that is why
nature chose them for spacetime. Even if there are extra hidden
dimensions, I think one could give a similar treatment and come to
similar conclusions.

There are at least two alternative pictures of spacetime foam, and I
have oscillated between them. One is the wormhole scenario
\cite{hwo,col}. Here the idea is that the path integral is dominated by
Euclidean spacetimes with large nearly flat regions (parent universes)
connected by wormholes or baby universes, though no good reason was
ever given as to why this should be the case. The idea was that one
wouldn't notice the wormholes directly, but only their indirect
effects. These would change the apparent values of coupling constants,
like the charge on an electron. There was an argument that the
apparent value of the cosmological constant should be exactly zero.
But the values of other coupling constants either were not determined
by the theory, or were determined in such a complicated way that there
was no hope of calculating them. Thus the wormhole picture would have
meant the end of the dream of finding a complete unified theory that
would predict everything.

A great attraction of the wormhole picture was that it seemed to
provide a mechanism for black holes to evaporate and disappear. One
could imagine that the particles that collapsed to form the black hole
went off through a wormhole to another universe or another region of
our own universe. Similarly, all the particles that were radiated from
the black hole during its evaporation could have come from another
universe, through the wormhole. This explanation of how black holes
could evaporate and disappear seems good at a hand waving level, but
it doesn't work quantitatively. In particular, one cannot get the
right relation between the size of the black hole and its
entropy. The nearest one can get is to say that the entropy of a
wormhole should be the same as that of the radiation-filled Friedmann
universe that is the analytic continuation of the wormhole. However,
this gives an entropy proportional to size to the three halves, rather
than size squared, as for black holes. Black hole thermodynamics is so
beautiful and fits together so well that it can't just be an accident
or a rough approximation. So I began to lose faith in the wormhole
picture as a description of spacetime foam.

Instead, I went back to an earlier idea \cite{hsf}, which I will refer
to as the quantum bubbles picture. Like the wormhole picture, this is
formulated in terms of Euclidean metrics. In the wormhole picture, one
considered metrics that were multiply connected by wormholes. Thus one
concentrated on metrics with large values of the first Betti number,
$B_1$. This is equal to the number of generators of infinite order in
the fundamental group. However, in the quantum bubbles picture, one
concentrates on spaces with large values of the second Betti number,
$B_2$. The spaces are generally taken to be simply connected, on the
grounds that any multiple connectedness is not an essential property
of the local geometry, and can be removed by going to a covering
space. This makes $B_1$ zero. By Poincare duality, the third Betti
number, $B_3$, is also zero. On this view, the essential topology of
spacetime is contained in the second homology group, $H_2$. The second
Betti number, $B_2$, is the number of two spheres in the space that
cannot be deformed into each other or shrunk to zero. It is also the
number of harmonic two forms, or Maxwell fields, that can exist on the
space. These harmonic forms can be divided into $B_{2+}$ self dual two
forms and $B_{2-}$ anti self dual forms. Then the Euler number and
signature are given by
$$\chi = B_{2+}+B_{2-}+2= \frac{1}{128 \pi^2} \int d^4 x \sqrt{g}
R_{\mu\nu\rho\sigma} R_{\alpha\beta\lambda\kappa}
\epsilon^{\mu\nu\alpha\beta}  \epsilon^{\rho\sigma\lambda\kappa},$$
$$\tau = B_{2+}-B_{2-}= \frac{1}{96 \pi^2} \int d^4 x \sqrt{g}
R_{\mu\nu\rho\sigma} R^{\mu\nu}\! _{\alpha\beta}
\epsilon^{\rho\sigma\alpha\beta}$$
if the spacetime manifold is compact. If it
is non compact, $\chi = B_{2+}+B_{2-}+1$ and the volume integrals
acquire surface terms.

Barring some pure mathematical details, it seems that the topology of
simply connected four manifolds can be essentially represented by
glueing together three elementary units, which I shall call
bubbles. The three elementary units are $S^2\times S^2$, $CP ^2$ and
$K3$. The latter two have orientation reversed versions, $\bar{CP}^2$
and $\bar{K3}$. Thus there are five building blocks for simply
connected four manifolds. Their values of the Euler number and
signature are shown in the table. To glue two manifolds together, one
removes a small ball from each manifold and identifies the boundaries
of the two balls. This gives the topological and differential
structure of the combined manifold, but they can have any metric.

\begin{table}
\centering
\begin{tabular}{|c||c|c|} \hline
  & Euler Number & Signature \\ \hline \hline
$S^2 \times S^2$ & 4 & 0  \\ \hline
$CP^2$ & 3 & 1 \\ \hline
$\bar{CP}^2$ & 3 & -1 \\ \hline
$K3$ & 24 & 16 \\ \hline
$\bar{K3}$ & 24 & -16 \\ \hline
\end{tabular}
\caption{ The Euler number and signature for the basic bubbles. }
\end{table}

If spacetime has a spin structure, which seems a physically reasonable
requirement, there can't be any $CP^2$ or $\bar{CP}^2$ bubbles. Thus
spacetime has to be made up just of $S^2\times S^2$, $K3$ and
$\bar{K3}$ bubbles. $K3$ and $\bar{K3}$ bubbles will contribute to
anomalies and helicity changing processes. However, their contribution
to the path integral will be suppressed because of the fermion zero
modes they contain, by the Atiyah-Singer index theorem. I shall
therefore concentrate my attention on the $S^2\times S^2$ bubbles.

When I first thought about $S^2\times S^2$ bubbles in the late 70s, I
felt that they ought to represent virtual black holes that would
appear and disappear in the vacuum as a result of quantum
fluctuations. However, I was never able to see how this correspondence
would work. That was one reason I temporarily switched to the wormhole
picture of spacetime foam. However, I now realize that my mistake was
to try to picture a single black hole appearing and
disappearing. Instead, I should have been thinking of black holes
appearing and disappearing in pairs, like other virtual
particles. Equivalently, one can think of a single black hole which is
moving on a closed loop. If you deform the loop into an oval, the
bottom part corresponds to the appearance of a pair of black holes and
the top, to their coming together and disappearing.

In the case of ordinary particles like the electron, the virtual loops
that occur in empty space can be made into real solutions by applying
an external electric field. There is a solution in Euclidean space
with an electron moving on a circle in a uniform electric field. If
one analytically continues this solution from the positive definite
Euclidean space to Lorentzian Minkowski space, one obtains an electron
and positron accelerating away from each other, pulled apart by the
electric field. If you cut the Euclidean solution in half along $\tau
=0$ and join it to the upper half of the Lorentzian solution, you get
a picture of the pair creation of electron-positron pairs in an
electric field. The electron and positron are really the same
particle. It tunnels through Euclidean space and emerges as a pair of
real particles in Minkowski space.

There is a corresponding solution that represents the pair creation of
charged black holes in an external electric or magnetic field. It was
discovered in 1976 by Ernst \cite{ernst} and has recently been
generalised to include a dilaton \cite{dgkt} and two gauge fields
\cite{2u1}. The Ernst solution represents two charged black holes
accelerating away from each other in a spacetime that is asymptotic to
the Melvin universe. This is the solution of the Einstein-Maxwell
equations that represents a uniform electric or magnetic field. Thus
the Ernst solution is the black hole analogue of the electron-positron
pair accelerating away from each other in Minkowski space. Like the
electron-positron solution, the Ernst solution can be analytically
continued to a Euclidean solution. One has to adjust the parameters of
the solution, like the mass and charge of the black holes, so that the
temperatures of the black hole and acceleration horizons are the same.
This allows one to remove the conical singularities and obtain a
complete Euclidean solution of the Einstein-Maxwell equations. The
topology of this solution is $S^2\times S^2$ minus a point which has
been sent to infinity.

The Ernst solution and its dilaton generalisations represent pair
creation of real black holes in a background field, as was first
pointed out by Gibbons \cite{gwg}.  There has been quite a lot of work
recently on this kind of pair creation.  However, in this paper I
shall be less concerned with real processes like pair creation, which
can occur only when there is an external field to provide the energy,
than with virtual processes that should occur even in the vacuum or
ground state. The analogy between pair creation of ordinary particles
and the Ernst solution indicates that the topology $S^2\times S^2$
minus a point corresponds to a black hole loop in a spacetime that is
asymptotic to $R^4$. But $S^2\times S^2$ minus a point is the
topological sum of the compact bubble $S^2\times S^2$ with the non
compact space $R^4$. Thus one can interpret the $S^2\times S^2$
bubbles in spacetime foam as virtual black hole loops. These black
holes need not carry electric or magnetic charges, and will not in
general be solutions of the field equations. But they will occur as
quantum fluctuations, even in the vacuum state.

If virtual black holes occur as vacuum fluctuations, one might expect
that particles could fall into them and re-emerge as different
particles, possibly with loss of quantum coherence. I have been
suggesting that this process should occur for some time, but I wasn't
sure how to show it. In fact Page, Pope and I did a calculation in
1979 of scattering in an $S^2\times S^2$ bubble, but we didn't know
how to interpret it \cite{bubble}.  I feel now, however, that I
understand better what is going on.

The usual semi-classical approximation involves perturbations about a
solution of the Euclidean field equations. One could consider particle
scattering in the Ernst solution. This would correspond to particles
falling into the black holes pair created by an electric or magnetic
field. The energy of the particles would then have to be radiated
again before the pair came back together again at the top of the loop
and annihilated each other. However, such calculations are unphysical
in two ways. First, the Ernst solution is not asymptotically flat,
because it tends to a uniform electric or magnetic field at infinity.
One might imagine that the solution describes a local region of field
in an asymptotically flat spacetime, but the field would not normally
extend far enough to make the black hole loop real. This would mean
that the field would have to curve the universe significantly. Second,
even if one had such a strong and far reaching field, it would
presumably decay because of the pair creation of real black holes.

Instead, the physically interesting problem is when a number of
particles with less than the Planck energy collide in a small region
that contains a virtual black hole loop. One might try and find a
Euclidean solution to describe this process. There are reasons to
believe that such solutions exist, but it would be very difficult to
find them exactly, and such effort wouldn't really be appropriate,
because one would expect the saddle point approximation to break down
at the Planck length. Instead, I shall take the view that $S^2\times
S^2$ bubbles occur as quantum fluctuations and that the low energy
particles that scatter off them have little effect on them. This
means one should consider all positive definite metrics on $S^2\times
S^2$, calculate the low energy scattering in them, and add up the
results, weighted with $\exp (-I)$ where $I$ is the action of the
bubble metric. If one were able to do this completely, one would have
calculated the full scattering amplitude, with all quantum
corrections. However, we neither know how to do the sum, nor how to
calculate the particle scattering in any but rather simple metrics.

Instead, I shall take the view that the scattering will depend on the
spin of the field and the scale of the metric on the bubble, but will
not be so sensitive to other details of the metric. In section 3 I
shall therefore consider a particular simple metric on $S^2\times
S^2$ in which one can solve the wave equations. I show that scattering
in this metric leads to a superscattering operator that does not
factorise. Hence there is loss of quantum coherence. In section 4, I
consider scattering on more general $S^2 \times S^2$ metrics, and
again find that the $\$$ operator doesn't factorise. The magnitude of
the loss of quantum coherence and its possible observational
consequences are discussed in section 5. Section 6 examines the
implications for the evaporation of macroscopic black holes, and
section 7 summarises the conclusions of the paper.
\section{The superscattering operator}

In this section I shall briefly describe the describe the results of
reference \cite{supers} on the superscattering operator $\$ $ which
maps initial density matrices to final density matrices,
$$ \rho _+^A{}_B=\$ ^A{}_{BC}^D \rho _-^C{}_D. $$
The idea is to define $n$ point expectation values for a field $\phi$
by a path integral over asymptotically Euclidean metrics,
$$ G_E (x_1, ..., x_n) = \prod_{j=1}^n
\left( - i \frac{\delta}{\delta J(x_j)} \right) Z[J] |_{J=0},$$
$$ Z[J] = \int d[\phi] e^{-I[\phi,J]}. $$
Because of the diffeomorphism gauge freedom, the expectation values
have an invariant meaning only in the asymptotic region near infinity,
where the metric can be taken to be that of flat Euclidean space. In
this region, one can analytically continue the expectation values to
points $x_1,~x_2,...,~x_n$ in Lorentzian spacetime. In Euclidean
space, the expectation values do not depend on the ordering, but in
Lorentzian space they do, because the field operators at timelike
separated points do not commute. In order to get the Lorentzian
Wightman functions $\langle \phi (x_1)\phi (x_2)...\phi (x_n)\rangle$,
one performs the analytical continuation from Euclidean space, keeping
a small positive imaginary time separation between the points $x_i$
and $x_{i-1}$. This generalises the usual Wick rotation from flat
Euclidean space to Minkowski space.

One can interpret the field operators $\phi $ in the Lorentzian flat
space near infinity as particle and
antiparticle annihilation and creation operators in the
usual way,
$$ \phi = \Sigma_i ( f_{i\pm} a_i + \bar{f}_{i\pm} b_i^\dagger), $$
$$ \phi^\dagger = \Sigma_i ( f_{i\pm} b_i + \bar{f}_{i\pm} a_i^\dagger), $$
where $\{f_{i\pm}\}$ are a complete orthonormal basis of
solutions of the wave equation that are positive frequency at future
or past infinity.

In the case of a black hole formed by gravitational collapse, the
initial states, which are $| \psi _i\rangle = I _i| 0\rangle $, where $I_i $
is a string of initial creation operators, form a complete basis
for the Hilbert space of fields on the background. However, the states
created by strings $F_i$ of creation operators at future infinity
don't form a complete basis, because one also has to specify the field
on the future horizon of the black hole. Indeed, it is this
incompleteness of states at future infinity that is responsible for
the radiation from the black hole. Spacetimes with closed loops of
black holes, like the Ernst solution, have both future and past black
hole horizons. Thus one might expect that in such spacetimes, the
states at both past and future infinity would fail to be a complete
basis for the Hilbert space.

If a spacetime is not asymptotically complete, that is, if the states
at future or past infinity are not a complete basis for the Hilbert
space, then quantum field theory on such a background will not be
unitary. We are used to this already. Quantum field theory on the
fixed background of a black hole formed by gravitational collapse
certainly is not unitary if one considers only the asymptotic states
at past and future infinity. It might be objected that such a
calculation ignores the back reaction of the particle creation and
that the final state consists not only of the asymptotic particle
states at future infinity, but also the black hole itself, which
contains the states needed to restore unitarity. The answer to the
first objection is that if one calculates the scattering on all
backgrounds and adds them up with the appropriate weights, one
automatically includes the back reaction. The answer to the second
objection is that with a closed loop of black holes, there is no black
hole in the final state: the black holes annihilate each other in a
way that is nonsingular at least in the Euclidean regime.

Even if the asymptotic states do not form a complete basis for the
Hilbert space, one can ask for the probability of observing the final
state $| \psi _3\rangle \langle \psi _4 | $ if one creates the initial
state $| \psi _1\rangle \langle \psi _2 | $ with strings of initial
creation operators. This will be related to
$$\langle I _2^{\dagger}F_3F_4^{\dagger}I _1\rangle. $$
If the asymptotic states at future and past infinity are complete
bases for the Hilbert space, this superscattering matrix element can
be factored,
$$\langle I _2^{\dagger}F_3F_4^{\dagger}I _1\rangle = \langle I
_2^{\dagger}F_3\rangle \langle F_4^{\dagger}I _1\rangle. $$
The second factor is the $S$ matrix and the first is its
adjoint. However, when black holes are present, the asymptotic states
are not complete and the $\$ $ operator does not factorise.

One can now use the Wightman functions to calculate the
superscattering operator. One can calculate
the expectation values of annihilation and creation operators by taking
the scalar products of the
Wightman functions with initial and final wave functions $f_i$ and
$\bar{f}_i$ on spacelike or null surfaces in the infinite future or
past. To get the right operator ordering, these surfaces should be
given small displacements in the imaginary time direction increasing
 from left to right in the expectation value.

\section{A simple bubble metric}

I now review a particularly simple example, previously discussed in
\cite{bubble}. Start with the four-sphere
$S^4$. This is conformally equivalent to flat Euclidean space $R^4$
with a point $p$ added at infinity. One can see this by blowing up the
round metric $g$ on $S^4$ with a conformal factor
$$\Omega = G(x,p),$$
 where $G$ is the Green function for the conformally invariant scalar
field.

Choose coordinates $\theta$, $\phi $, $\chi $ and $\psi $ on the
four-sphere. Now identify the point with coordinates $(\theta , \phi ,
\chi , \psi )$ with the point $(\pi -\theta , -\phi , \pi -\chi , \pi
-\psi )$. This identification has two fixed points $q$ and $r$ at
opposite points on the equator $\psi =\pi /2$. At the fixed points,
the identified sphere is an orbifold, not a manifold. However, one can
make it a manifold again by cutting out small neighbourhoods of the
two fixed points and replacing them by an Eguchi Hanson metric and an
Eguchi Hanson with the opposite orientation respectively. This
identification and surgery changes the topology of the $S^4 $ into
$S^2\times S^2$. One can now pick a point $p$ which is neither $q$
nor $r$, and send it to infinity with a conformal factor
$$\Omega (x)= G(x,p).$$
This gives an asymptotically Euclidean metric with topology $S^2\times
S^2-\{p\}$.

There will be well-defined expectation values or Green functions on
this Euclidean space, which one can construct with image charges. One
can then use these expectation values to calculate particle scattering
by the bubble. One can define the data for ingoing and outgoing
plane waves on the light cone of the infinity point $p$, on which the
metric is asymptotically Lorentzian.  This light cone is like $\cal
I^-$ and $\cal I^+$ in asymptotically flat space. One then uses the
analytically continued expectation values to propagate the in states
to the out states.

This scattering calculation was done some time ago but it was not
understood how to interpret it. I now think I see what is happening.
A positive frequency solution of the wave equation in Minkowski space
can be analytically continued to be a solution that is holomorphic on
the lower half of Euclidean space. One can conformally map Euclidean
space to $S^4-\{p\}$ so that $p$, $q$ and $r$ lie on the equator. Then
a positive frequency solution is holomorphic on the lower half
sphere. The identification I described connects points in the lower
half sphere with points in the upper half sphere. Thus, it maps a
positive frequency function into a negative frequency one.

\begin{figure}
\leavevmode
\centering
\epsfysize=10cm
\epsfbox{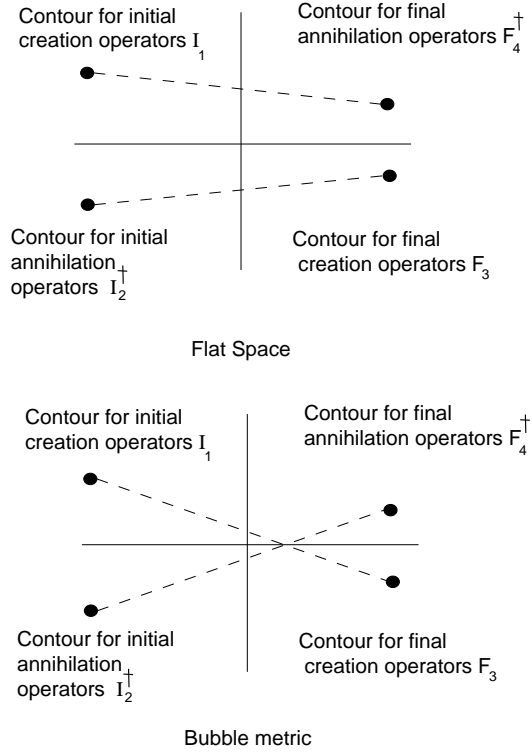}
\caption{The complex $t$ plane for the calculations of the
  superscattering operator, showing how the Wightman functions are
  integrated over contours with a small imaginary time displacement.
  The upper diagram corresponds to flat space and the lower to the
  extra scalar products that occur in the identified sphere bubble.
\label{fig1}}
\end{figure}

Recall from section 2 that the $\$ $ operator element
$$\langle I _2^{\dagger}F_3F_4^{\dagger}I _1\rangle $$
can be calculated by taking the scalar product of the Wightman
functions with the initial and final wave functions on $\cal I^-$ and
$\cal I^+$. To get the right operator ordering, the contours of
integration over the affine parameter $u$ on the null geodesic
generators of $\cal I^-$ and $\cal I^+$ should be displaced slightly
in the order $2341$ in increasing imaginary $u$. If there were no
identifications and the spacetime was just flat space, the negative
frequency wave from the final state annihilation operators
$F_4^{\dagger}$ can only propagate upwards in the complex $u$ plane.
This means the only contour with which they can have a nonzero scalar
product is that for the initial creation operators $I_1$. Similarly,
the positive frequencies from the final state creation operators $F_3$
can only propagate downwards in imaginary $u$, and can have a nonzero
scalar product only with the contour on which the initial state
annihilation operators $I_2^{\dagger}$ act
(Figure \ref{fig1}).   Thus in this case the $\$
$ operator factorises,
$$\langle I _2^{\dagger}F_3F_4^{\dagger}I _1\rangle = \langle I
_2^{\dagger}F_3\rangle \langle F_4^{\dagger}I _1\rangle. $$
There is a unitary evolution with no loss of quantum coherence.

On the identified four sphere however, the data from the final state
annihilation operators $F_4^{\dagger}$ will also propagate downwards
 from an image of the contour $4$ below the real $u$ axis. It thus can
have a nonzero scalar product with the contour $2$ on which the
initial state annihilation operators $I_2^{\dagger}$ act. Similarly,
there can be a non zero scalar product between the data from the final
state creation operators $F_3$ and the initial state creation
operators $I_1$ (Figure \ref{fig1}).

These scalar products have been calculated for conformally invariant
fields of spin $s$ propagating on this background. For each particle
with initial and final momenta $k_2^{\mu}$ and $k_4^{\mu}$, the
$4\rightarrow 2$ scalar product gives a factor
$$ \langle k_4 | k_2 \rangle = -\frac{q^2}{8\pi} e^{ip \cdot (k_2 +
  k_4)} J_{2s}  \left( \left[- \frac{1}{2} k_2 \cdot k_4 q^2 + (q \cdot k_2)
(q \cdot k_4) \right]^{1/2} \right) ,$$
where $p^{\mu} ={1\over 2}(x_q^{\mu}+x_r^{\mu})$ and
$q^{\mu}=x_q^{\mu}- x_r^{\mu}$. The scalar product $3\rightarrow 1 $
has a similar factor for each particle, but $k_1^{\mu}$ and $k_3^{\mu}$
appear with the opposite signs.

There will be factors like this for each of the $n$ particle lines
passing through the bubble.  There will also be a factor $\Delta
^{-1/2}\exp (- I) $ where $\Delta $ is the determinant of the
conformally invariant field wave operator and $I ={3 \over 8}\pi q^2$ is
the action of the asymptotically Euclidean bubble metric.  One them
integrates over the positions of the points $q$ and $r$ or
equivalently over the vectors $p$ and $q$.  The integral over all $p$
produces $\delta (k_2+k_4-k_3-k_1)$.  This does not guarantee energy
momentum conservation because it would be satisfied by $k_1=k_2 \neq
k_3=k_4$.  As discussed below, energy momentum conservation comes from
the path integral over all metrics equivalent under diffeomorphisms.
The integral over all $q$ averages over the orientation and scale of
the bubble metric.  The dominant contribution to the integral over the
scale will come from bubbles of order the Planck size.

These nonzero scalar products that would not occur in flat space have
two consequences. First, consider a field $\phi$ with a global
symmetry such as $U(1)$ that is not coupled to a gauge field. Take the
initial state operators $I_1$ and $I_2$ to be particle creation
operators and the final state operators $F_3$ and $F_4$ to be
anti-particle creation operators. Then there will be a nonzero
probability for a particle to change into its anti-particle. This is
what one would expect. In the presence of black holes, real or
virtual, global charges will not be conserved. However, if the
particles are coupled to a gauge field, averaging over gauges will
make the amplitude zero unless the gauge charge is conserved.
Similarly, averaging over diffeomorphisms, the gravitational gauge
degrees of freedom, should ensure that the amplitude is zero unless
energy is conserved.  As was seen above, energy conservation is not
guaranteed by integration over the position of the bubble. When there
is loss of quantum coherence, it is only local symmetries and not
global ones that imply conservation laws.

The second consequence of the nonzero scalar products is that the $\$
$ operator giving the probability to go from initial to final will not
factorise into an $S$ matrix times its adjoint. This means that the
evolution from initial to final will be non-unitary and will exhibit
loss of quantum coherence. This is what you might expect in a bubble
with non-trivial topology, because the Euler number of three will mean
that one cannot foliate the spacetime with a family of time
surfaces. One thus cannot show there is a unitary Hamiltonian
evolution. However, any suggestion that quantum coherence may be lost
seems to arouse furious opposition. It is almost like I was attacking
the existence of the ether.

\section{Scattering by black hole loops}

The metric considered in the last section was a special limiting case
of an asymptotically Euclidean $S^2\times S^2-\{p\}$ metric. However,
one might be concerned that because it so special, scattering in it
would not be typical of $S^2\times S^2-\{p\}$ bubbles. In this
section I shall therefore consider scattering in a different class of
metrics that correspond more directly with the intuitive picture of
$S^2\times S^2$ bubbles as closed loops of real or virtual black
holes.

One cannot analytically continue a general real Euclidean metric to a
section of the complexified manifold on which the manifold is real and
Lorentzian. This does not matter for scattering calculations, because
one can analytically continue to Lorentzian at infinity, and one
does not directly measure the metric at interior points, but one
integrates over all possible metrics. The idea is that the path
integral over all Lorentzian metrics is equivalent to a path integral
over all Euclidean ones in a contour integral sense. However, in order
to give the scattering a physical interpretation, it is helpful to
consider metrics that have both Euclidean and Lorentzian
sections. This will be guaranteed if the metric has a hypersurface
orthogonal Killing vector. If the metric is asymptotically Euclidean,
one can interpret this Killing vector as corresponding to a Lorentz
boost at infinity. For simplicity, I shall also assume that there is a
second commuting hypersurface orthogonal Killing vector corresponding
to rotations about an axis. The is the maximum symmetry that an
asymptotically Euclidean metric on $S^2\times S^2-\{p\}$ can have. In
particular, virtual black holes can not be spherically symmetric.

The Lorentzian section of the metric will have a structure like that of
the $C$ metric or the Ernst solution, with two black holes accelerating
away from each other. By the positive action theorem, there are no
asymptotically Euclidean solutions of the vacuum Einstein equations
with topology $S^2\times S^2-\{p\}$. The $C$ metric has singularities on
the axis, which can be interpreted as cosmic strings pulling the black
holes apart, and the Ernst solution is asymptotic not to flat Euclidean
space, but to the Euclidean Melvin solution. However, as was said
earlier, I shall consider asymptotically Euclidean metrics on
$S^2\times S^2-\{p\}$ that correspond not to real black holes, but to
virtual black hole loops that arise as vacuum fluctuations. These will
not be solutions of the Einstein equations, and will be similar to the
$C$ metrics, but without singularities on the axis.

\begin{figure}
\leavevmode
\centering
\epsfysize=10cm
\epsffile{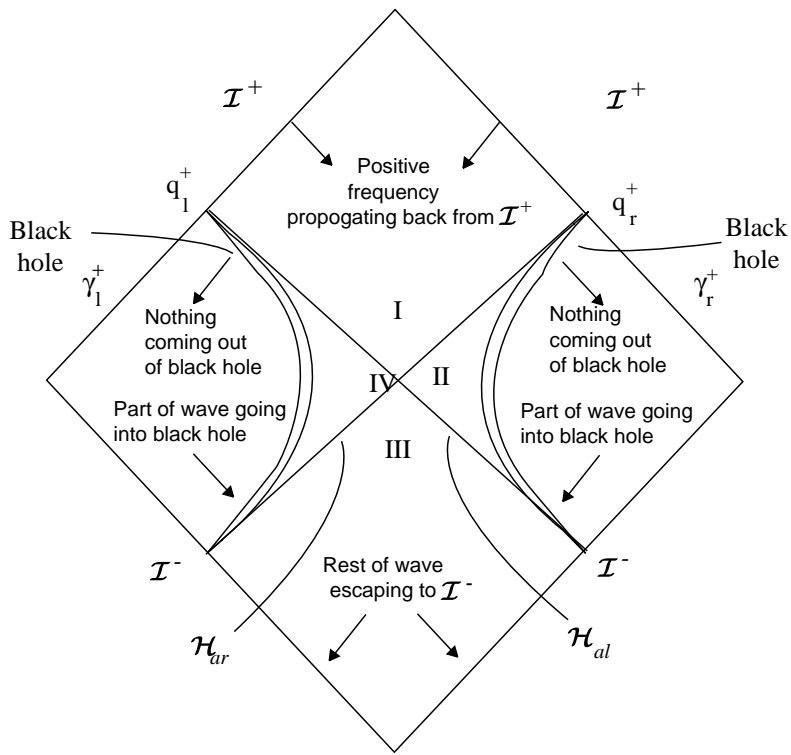}
\caption{The Lorentzian section of an asymptotically Euclidean metric
  on $S^2 \times S^2 - {pt}$.
\label{fig2}}
\end{figure}

The Lorentzian metrics will be asymptotically flat with zero
mass.\footnote{ Lorentzian solutions with non zero mass have a weak
  conformal singularity at the infinity point.  However I shall ignore
  this for center of mass energies low compared to the Planck mass.
  Such a singularity would affect the propagation only in the
  asymptotic region.} They will have good past and future null
infinities $\cal I^-$ and $\cal I^+$, which are the light cones of the
point $p$ at infinity in the conformally compactified Euclidean metric
or the spatial infinity point $I^0$ in the conformally compactified
Lorentzian metric. The boost Killing vector $\xi $ and the
axisymmetric Killing vector $\eta $ can be extended to $\cal I^{\pm}$.
On $\cal I^+$, $\xi $ will have two fixed points, $q_l^+$ and $q_r^+$,
on the left and right of figure \ref{fig2}. The past light cones of
these fixed points, apart from the two generators $\gamma _l^+$ and
$\gamma _r^+$, which lie in $\cal I ^+$, form the left and right
acceleration horizons ${\cal H}_{al}$ and ${\cal H}_{ar}$. These light
cones focus again to two fixed points $q_r^-$ and $q_l^-$ on the right
and left of $\cal I^-$ respectively. The acceleration horizons divide
the region outside the black holes into the left and right Rindler
wedge, labelled IV and II, and the future and past regions, labelled I
and III.

There are also two black hole horizons ${\cal H}_{bl}$ and ${\cal H}
_{br} $.  The horizon ${\cal H}_{bl}$ consists of ${\cal H}_{bl}^+$,
the future horizon of the left black hole, and ${\cal H}_{bl}^-$, the
past horizon of the right black hole. Similarly, ${\cal H}_{br}$
consists of the future horizon of the right black hole and the past
horizon of the left black hole.

The region outside the black holes is globally hyperbolic. One can
therefore analyse the behaviour of a massless field $\phi$ in a manner
similar to that on static black holes \cite{hbh}. One can take a past
Cauchy surface to be $\cal I ^-$ and the past left and right black
hole horizons ${\cal H}_{br}^-$ and ${\cal H}_{bl}^- $.  Similarly,
$\cal I ^+$ and future black hole horizons will form a future Cauchy
surface. Now consider a solution $\mu $ of the wave equation which has
positive frequency on $\cal I ^+$ with respect to the affine parameter
and zero data on on the future black hole horizons.  As Yi \cite{yi}
has pointed out, it is reasonable to ignore $\gamma _l^+$ and $\gamma
_r^+$ as sets of measure zero on $\cal I ^+ $, and to take the support
of $\mu $ to be away from them. In other words, one ignores waves
directed exactly along the axis asymptotically.

In this case, $\mu $ will propagate backwards through the future
region I to the future V formed by the future halves of the
acceleration horizons. On ${\cal H}_{a l}^+$ and ${\cal H} _{ar}^+ $ one
can decompose $\mu $ into modes with definite frequency $\omega ' $
with respect to the Rindler time associated with the boost Killing
vector $\xi $. One can also separate into eigenmodes with respect to
the axial Killing vector $\eta $, but the wave equation probably
cannot be separated in the remaining two dimensions. I shall therefore
label the eigenmodes $\psi _{\omega ' mn} $ where $n $ labels the
eigenmodes of the wave equation in the remaining two dimensions.

One can now consider the wave equation in the right hand Rindler wedge
II. Since one is ignoring $\gamma _r^+$ as a set of measure zero, a
Cauchy surface for this region will be the future acceleration horizon
${\cal H}_{ar}^+$ and the future black hole horizon ${\cal
H}_{br}^+$. The data for $\mu $ will be zero on ${\cal H}_{br}^+$ (by
assumption), and will be a mixture of eigenmodes $\psi _{\omega ' mn}$
on ${\cal H}_{ar}^+$. A fraction $\Gamma_{\omega' m n} $ of
the flux of  each eigenmode will cross
the past black hole horizon ${\cal H}_{bl}^-$, and the remaining
$(1-\Gamma_{\omega' mn} )$ will reflect on the effective potential and will
cross
the past acceleration horizon ${\cal H}_{a l}^-$. Similarly, one can
solve the wave equation in the left hand Rindler wedge IV and find
that a fraction $\Gamma_{\omega' mn} $ goes into the black hole and a fraction
$(1-\Gamma_{\omega' mn} )$ crosses the past acceleration horizon.

\begin{figure}
\leavevmode
\centering
\epsfysize=10cm
\epsffile{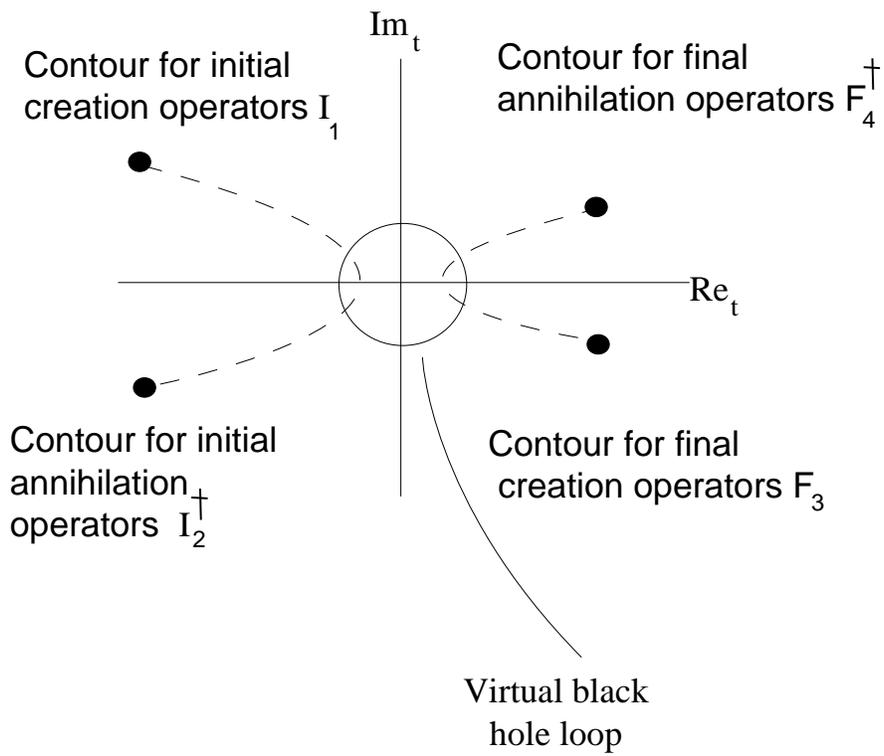}
\caption{The complex $t$ plane for scattering on an asymptotically
  Euclidean virtual black hole metric.
\label{fig3}}
\end{figure}

One now has data on the two past acceleration horizons ${\cal
  H}_{al}^-$ and ${\cal H} _{ar} ^-$ and can solve the wave equation
on the past region III. For each eigenmode, the data on the left and
right acceleration horizons will both be reduced by the same factor
$(1-\Gamma_{\omega' mn} ) ^{1/2}$. Thus it seems likely that $\mu $
will be purely positive frequency on $\cal I ^-$. However, it will not
be purely positive frequency on the black hole horizons, because it is
non zero on the past parts ${\cal H}_{br}^-$ and ${\cal H} _{bl} ^-$
but it is zero by assumption on the future parts ${\cal H}_{bl}^+$ and
${\cal H} _{br} ^+$.  This means that an observer at $\cal I ^+$ will
observe particles in the mode $\mu $, contrary to the claims of Yi
\cite{yi}.  Another way of saying this is that the positive
frequencies from the final state creation operators $F_3$ on $\cal I
^+$ will have a non zero scalar product with the final state
annihilation operators $F_4^{\dagger}$, so that
$$\langle F_3F_4^{\dagger}\rangle \neq 0.$$ Similarly, the positive
frequencies from the initial creation operators $I_1$ can go into the
black holes and have a non zero scalar product with the initial
annihilation operators.  This gives a diagram like Figure \ref{fig3}.
Note that the initial annihilation and creation operators can belong
to different particle species from those of the final operators.  This
is what one might expect because the No Hair theorems imply that a
black hole forgets what fell into it apart from charges coupled to
gauge fields.  It means that the full superscattering matrix element
will not factorise. Further discussion of scattering in metrics of
this type will be given in another paper.

\section{Observational consequences}

Obviously, quantum coherence is not lost under normal conditions to a
very high degree of approximation, so one has to ask what order of
magnitude the bubble scattering calculation would indicate. It seems
that the scattering at low energies depends strongly on the spin of
the field. One can see this explicitly in the case of the identified
sphere metric in section 3. Here the amplitudes were products of
Bessel functions $J_{2s}(c)$ for each pair of momenta, where $s$ was
the spin and $c $ was a quantity of order of the center of mass energy
in the scattering. For low energy scatterings, $c \ll 1$,
$$J_{2s}(c)\approx c^{2s}.$$
These amplitudes are  of the same order as those that would be
produced by effective interactions of the form
$$m_p^{4-2n(1+s)}(\phi )^{2n},$$
where $n\ge 2$ is the number of pairs
of ingoing or outgoing momenta scattered through the bubble. The field
$\phi $ in the effective interaction is the scalar field for $s=0$ and
the spinor field for $s={1\over 2}$. For $s=1$, it is the field
strength $F_{\mu \nu}$. The scattering calculations have not been done
explicitly for spin $3\over 2$ and $2$, but on this basis one would
expect the effective interactions to involve the gradient of the spin
$3\over 2$ field and the curvature respectively.

One would like to know whether this spin dependence of the scattering
is peculiar to the special bubble metric considered in section 3, or
whether it is a general feature. In fact, consideration of scattering
in the more general metrics of section 4 suggests that the effective
interactions depend on spin in a similar way. The non-factoring part
of the scattering can be thought of as a scattering cross section for
a wave to get into a black hole and a thermal factor. Calculations of
scattering by static black holes indicate that for black holes much
smaller than the wave length $\omega ^{-1}$, the absorption cross
sections are of the order of the geometrical cross sections for both
$s=0$ and $s={1\over 2}$, while they are of order $\omega ^{2}$ for
$s=1$. The Bose-Einstein thermal factor will be of order $\omega
^{-1}$ while the Fermi-Dirac factor will be order $1$. Thus one will
get the same $c^{2s}$ dependence on spin. It is therefore reasonable
to suppose that any bubble metric will give effective interactions of
the same order.

The effective interactions induced by bubbles are local, in that the
scale of the bubble will be of order the Planck length, while the
center of mass wavelength will be larger for low energy
scatterings. However, they will be nonlocal in that they will mix up
the separation that one has in flat space between the diagrams for the
$S$ matrix and its adjoint. This separation is of order $\epsilon $,
and one takes the limit $\epsilon \rightarrow 0$. Thus the separation
will become less than the size of the bubble. If the effective
interactions had been purely local, they would have produced a unitary
evolution, but the fact that they are nonlocal means that quantum
coherence can be lost \cite{bps,uw}.

One can see that almost all these effective interactions are
suppressed by factors of the Planck mass. The only exceptions are
scalar fields, which would get an effective $\phi ^4$ or $\phi^2\phi
^2$ interaction, with coefficients of order one. But we have never yet
observed an elementary scalar field. Particles like the pion are
really bound states of fermions. When scattering off a bubble, they
would behave like individual fermions. This suggests that we may never
observe the Higgs particle, because it will be strongly coupled to
every other scalar field, or that if we do detect it, it will turn out
to be a bound state of fermions.

Effective interactions between fermions will be suppressed by two
powers of the Planck mass for a four fermion vertex and five powers
for a six fermion vertex. The first could contribute to $K_L^0$ decay
and the second to baryon decay. However, the lifetimes are of the
order of $10^7$ and $10^{64}$ years respectively, so they are not of
much experimental interest. The quantum coherence violating effective
interactions induced between spin $1$ fields are even more suppressed,
so we wouldn't have observed them. On the other hand, there might be a
$\psi ^2\phi ^2$ fermion scalar effective interaction that was
suppressed by only one power of the Planck mass. The possible
consequences of such an interaction will be investigated elsewhere.

Another observational feature that might be explained by loss of
quantum coherence is the fact that the $\theta $ angle of QCD is
zero. One way of interpreting the $\theta $ angle is to regard the QCD
vacuum as a coherent sum
$$\sum e^{i\theta} | n\rangle $$
of states labelled with a winding number $n$. Although there are no
asymptotically Euclidean vacuum solutions, there are asymptotically
Euclidean Einstein-Maxwell solutions. These have an asymptotically
self dual uniform Maxwell field at infinity. They were investigated by
one of my students, Alan Yuille, and are in his PhD thesis, but are
otherwise unpublished. If one takes a $U(1)$ subgroup of a Yang-Mills
group, one can promote them to Einstein-Yang-Mills solutions. The
ordinary Yang-Mills instantons in flat space have self dual Yang-Mills
fields which can be taken to be uniform over sufficiently small
regions. Thus one could imagine glueing small bubbles on to a flat
space Yang-Mills instanton and obtaining an instanton with warts that
was a solution of the field equations. One might expect that the
bubbles, or warts, would produce loss of coherence between the
different $| n\rangle $ vacua. In other words, there would be a
nonzero probability to go from the product density matrix
$$\sum e^{i(n-m)\theta } | n \rangle \langle m | $$
to a  density matrix with other coefficients. Presumably the density
matrix would tend to the state with lowest energy, which is probably
the $\theta =0$ density matrix with equal coefficients.

If $\theta $ were nonzero (and in flat space Yang-Mills theory, there
is no reason why it shouldn't be), it would have produced effects like
a dipole moment for the neutron, which would have been observed. To
explain the absence of a dipole moment, the Peccei-Quinn \cite{pq}
mechanism was proposed. The original version of the mechanism was
ruled out because it predicted an axion of a few hundred KeV mass that
was not observed. There was a grand unified theory version of the
mechanism, which would have given rise to a very light and weakly
interacting axion. At one time, it was hoped that this axion might
make up the cold dark matter required to give the universe the
critical density. However, recent work on the damping of axion cosmic
strings \cite{bs} has almost closed the window of possible masses for
the axion.  So we badly need an explanation of the zero dipole moment
of the neutron. My bet is that it is loss of quantum coherence.

In the case of the wormhole picture, it seemed at first that quantum
coherence would be lost because wormholes would connect the upper and
lower halves of diagrams for the $\$ $ matrix.  However, it turned out
that that effects of wormholes on low energy physics could be
described by a number of alpha parameters \cite{col}. These would act
as the coupling constants for ordinary local effective interactions
that didn't lose quantum coherence. Their values wouldn't be
determined by the theory. However, one could conduct experiments to
measure all the effective coupling constants up to a certain order.
After that, there would be no unpredictability or loss of quantum
coherence. One would have ordinary quantum field theory with coupling
constants that couldn't be predicted but could be chosen to agree with
experiment.

Could the situation be similar with the quantum bubbles picture?
Could the unpredictability associated with loss of quantum coherence
be absorbed into a lack of knowledge of coupling constants?  I can't
rule this out, but I don't think it will be the case. There is an
important difference between the wormhole and bubble pictures. With
wormholes, one can integrate over the position of each end of the
wormhole separately. This allows the effect of the wormhole to be
factorised into separate local interactions at each end of the
wormhole. However, with a quantum bubble, there is only one integral
over the position of the bubble. Thus, one cannot factorise the effect
of a bubble. It will therefore give rise to a nonlocal interaction
that connects the evolution of a quantum state with that of its
complex conjugate. I therefore expect that when one sums over all the
bubbles in spacetime foam, one will still get loss of quantum
coherence.

\section{Evaporation of macroscopic black holes}

The picture of virtual black holes as occurring in pairs and
corresponding to $S^2\times S^2$ topological fluctuations has
implications for the end point of the evaporation of a macroscopic
black hole. For twenty years, I tried to think of a Euclidean geometry
that would describe the disappearance of a single black hole. But the
only thing that seemed possible was a wormhole, and I have already
said why I came to reject that idea. However, I now think that when a
black hole evaporates down to the Planck size, it won't have any
energy or charge left, and it will just disappear into the sea of
virtual black holes. If this picture is correct, it implies that two
dimensional models can't describe the disappearance of black holes in
a way that is nonsingular. This agrees with our experience. The best
we can do in two dimensions is the RST \cite{rst} model. In this, a
black hole evaporates down to zero mass. However, one then has to cut
the solution off by hand and join on the vacuum solution. This is very
ad hoc and introduces a naked singularity. Strominger and Polchinski
\cite{sp} have tried to argue that a baby universe branches off.
However, I think that is wrong for the reasons for which I rejected
the wormhole scenario.

\section{Conclusions}

It seems that topological fluctuations on the Planck scale should give
spacetime a foam-like structure. The wormhole scenario and the quantum
 bubbles picture are two forms this foam might take. They are
characterized by very large values of the first and second Betti
numbers respectively. I argued that the wormhole picture didn't really
fit with what we know of black holes. On the other hand, pair creation
of black holes in a magnetic field or in cosmology is described by
instantons with topology $S^2\times S^2$. This shows that one can
interpret $S^2\times S^2$ topological fluctuations as closed loops of
virtual black holes.

I then went on to discuss particle scattering by $S^2\times S^2$
bubbles.  Because of the non-trivial topology, one cannot cover the
manifold with a family of time surfaces. One cannot therefore act with
a Hamiltonian and get a unitary evolution from the initial state to
the final one. It is therefore possible that quantum coherence could
be lost, and I showed that indeed it was, both explicitly, in a simple
bubble metric, and in more general cases. I gave estimates of the
magnitude of bubble induced effects. They are all suppressed by powers
of the Planck mass, with the exception of scalar fields. We have not
yet observed an elementary scalar particle, and I predict we never
will. Another prediction of the quantum bubbles picture is that the
$\theta $ angle of QCD should be exactly zero, without having to
invoke the existence of an axion. This is almost ruled out by
observation, anyway. There may well be other predictions of the
quantum bubble picture which are testable at low energy. Thus, the
question of the Planck scale structure of spacetime may not be as
esoteric as it is sometimes made out to be. In the fluctuations in the
microwave background, we are already observing effects on scales of
about $10^6$ Planck lengths.  This would have been the horizon size of the
universe at the time the fluctuations were produced during
inflation. So quantum gravity is real physics. I think it is quite
possible that we can observe the consequences of spacetime structure
on even smaller scales. This will be one of the challenges for the
next few years. Unless quantum gravity can make contact with
observation, it will become as academic as arguments about how many
angels can dance on the head of a pin.


\begin{thebibliography}{19}
\bibitem{hwo}S. W. Hawking, Phys. Rev. D {\bf 37}, 904 (1988).
\bibitem{col}S. Coleman, Nucl. Phys. {\bf B130}, 643 (1988).
\bibitem{hsf}S.  W.  Hawking, Nucl. Phys. {\bf B144}, 349 (1978).
\bibitem{ernst}F. J. Ernst, J. Math. Phys. {\bf 17}, 515 (1976).
\bibitem{dgkt}H.F. Dowker, J.P.  Gauntlett, D.A. Kastor and J.
  Traschen, Phys. Rev.  D {\bf 49}, 2909 (1994).
\bibitem{2u1}S.F. Ross, Phys. Rev. D {\bf 49}, 6599 (1994).
\bibitem{gwg}G.W. Gibbons, in {\it Fields and Geometry 1986},
  Proceedings of the 22nd Karpacz Winter School of Theoretical
  Physics, Karpacz, Poland, edited by A. Jadczyk (World Scientific,
  Singapore, 1986).
\bibitem{bubble}S.W. Hawking, D.N. Page and C.N. Pope,
  Nucl. Phys. {\bf B170}, 283 (1980).
\bibitem{supers}S.W. Hawking, Commun. Math. Phys. {\bf 87}, 395
  (1982).
\bibitem{hbh}S. W. Hawking, Commun. Math. Phys. {\bf 43}, 199 (1975).
\bibitem{yi}P. Yi, Phys. Rev. Lett. {\bf 75}, 382 (1995); P. Yi,
  ``Quantum stability of accelerated black holes'', preprint
  CU-TP-690, hep-th/9505021.
\bibitem{bps}T. Banks, L. Susskind and M.E. Peskin, Nucl. Phys. {\bf
    B244}, 125 (1984).
\bibitem{uw}W.G. Unruh and R.M. Wald, Phys. Rev. D {\bf 52}, 2176
  (1995).
\bibitem{pq}R.D. Peccei and H.R. Quinn, Phys. Rev. Lett. {\bf 38},
  1440 (1977).
\bibitem{bs}R.A. Battye and E.P.S. Shellard, Phys. Rev. Lett. {\bf 73},
  2954 (1994).
\bibitem{rst}J.G. Russo, L. Susskind and L. Thorlacius,
  Phys. Rev. D {\bf 46}, 3444 (1992).
\bibitem{sp}J. Polchinski and A. Strominger, Phys. Rev. D {\bf 50},
  7403 (1994).
\end{thebibliography}
\end{document}